\documentclass[%
 reprint,
 amsmath,amssymb,
 aps,
]{revtex4-2}

\usepackage{graphicx}% Include figure files
\usepackage{xcolor}
\usepackage{dcolumn}% Align table columns on decimal point
\usepackage{bm}% bold math
\begin{document}

\preprint{APS/123-QED}

\title{Cavity quantum electrodynamics on a nanocapillay fiber using a composite photonic crystal symmetric cavity}

\author{Srinu Gadde}
\author{Jelba John}
\author{Elaganuru Bashaiah}
\author{Ramachandrarao Yalla}
\email{rrysp@uohyd.ac.in}
 
 \affiliation{School of Physics, University of Hyderabad, Hyderabad, Telangana 500046, India} 
\date{\today}

\begin{abstract}
We report the cavity quantum electrodynamics conditions using a composite photonic crystal symmetric cavity (CPCSC) on an optical nanocapillary fiber (NCF). The CPCSC is formed by combining the NCF and symmetric defect mode nano-grating. The maximum channeling efficiency up to $87$\% is realized by placing the single quantum emitter at the anti-node position of the CPCSC. The current platform paves the way for manipulating single photons with applications in quantum technology.
\end{abstract}

\maketitle

%\tableofcontents

\section{INTRODUCTION}
The enhancement of light-matter interaction (LMI) by creating a cavity is generally known as cavity quantum electrodynamics (QED) \cite{hung2013trapped, thompson2013coupling}. The cavity-QED has numerous applications in quantum technologies and quantum internet \cite{kimble1998strong, vahala2003optical, yoshie2004vacuum, hennessy2007quantum, ritter2012elementary, reiserer2015cavity}. To achieve cavity-QED conditions through the Purcell enhancement,  various types of nanowaveguide-based cavities have proved to be promising candidates with recent implementations \cite{hausmann2013coupling,goban2014atom,nayak2014optical,nayak2018nanofiber}. Among them, optical nanofiber (ONF) based cavities have proven to be a promising candidate for the efficient collection and manipulation of single photons in a controlled manner due to their ability to independently confine optical fields in the transverse and longitudinal directions with easy integration with single-mode fibers (SMF) for applications in quantum technologies and quantum networks \cite{nayak2018nanofiber}. They are advantageous since they offer tight confinement of optical fields in both directions, enabling enhanced interaction strength between guided modes and surrounding media. They can serve as transmission lines between quantum nodes in a quantum network \cite{kimble2008quantum}. Nevertheless, it is still challenging to effectively channel single photons from a single quantum emitter (SQE) into a particular mode \cite{bremer2022fiber}. 

Various types of cavity structures such as photonic crystal (PhC) and fiber Bragg grating (FBG) have been proposed, designed and experimentally demonstrated in the case of ONFs \cite{sadgrove2013photonic, yalla2014cavity, schell2015highly, kato2015strong, keloth2015diameter, keloth2017fabrication, li2017optical, li2018tailoring, nayak2019real, qing2019simple, takashima2019fabrication, yang2020photonic}. The essential role of formed cavity structures is used to enhance the spontaneous emission rates of  SQE \cite{nayak2018nanofiber}. The PhC/FBG cavities on ONFs have many applications, such as total spontaneous emission from atom can be coupled to guided modes of ONF \cite{le2009cavity}, tunability of cavity' resonance wavelength \cite{schell2015highly, yalla2020design}, measurement of ONF's diameter \cite{keloth2015diameter}, study of photothermal properties of cavity \cite{wang2019photothermal}, optical sensing applications at nanoscales \cite{yang2020photonic}, and efficient collection and channeling of single photons through one-sided guided modes for single photons source based quantum applications \cite{yalla2022one}. 

The experimental realization of PhC/FBG cavities on tapered ONFs can be broadly classified into two categories. The first is creating a PhC/FBG cavity on ONF using femtosecond laser ablation and focused ion beam milling techniques. The recent demonstrations that include periodic index modulating nanogrooves \cite{thompson2013coupling, nayak2018nanofiber, schell2015highly, takashima2016detailed, takashima2019fabrication}, nano-craters \cite{keloth2017fabrication,le2009cavity} and rectangular holes \cite{li2017optical, li2018tailoring, sahu2022optimization}. The second is the composite PhC cavity (CPCC) \cite{sadgrove2013photonic, yalla2014cavity, keloth2015diameter, yalla2022one}. The CPCC is formed by combining the external defect mode nano-grating (DMG) with the waist region of ONF. Depending upon the geometry of the nano-grating around the defect center, CPCC is further classified into two categories. One is the composite PhC symmetric cavity (CPCSC) \cite{yalla2014cavity, keloth2015diameter, yalla2020design}, in which the nano-grating is symmetric about the defect center. Using the symmetric CPCC technique, the enhanced spontaneous emission from quantum dots leads to a maximum channeling efficiency ($\eta$) of $65$\% through the guided modes of SMF was demonstrated \cite{yalla2014cavity}. In this case, the randomly polarized quantum dot is placed on the ONF surface, which limits the LMI. The other composite PhC asymmetric cavity, in which the nano-grating is asymmetric about the defect center. The formation of the one-sided cavity on the ONF has been experimentally demonstrated \cite{yalla2022one}.

Very recently, for the efficient collection of single photons from SQE and to further enhance the LMI, a new type of hollow core fiber termed a capillary fiber, whose inner diameter is in the sub-micron range, was proposed and experimentally demonstrated \cite{faez2014coherent}. However, in these efforts, $\eta$-value of 18\% was demonstrated. The measured value is a low $\eta$-value due to a thicker outer diameter, resulting in weak field confinement within the capillary fiber. Furthermore, our group proposed and reported a maximum $\eta$-value of $52\%$ by placing a radially polarized SQE inside the optical nanocapillary fiber (NCF), whose inner and outer diameters are in the subwavelength range \cite{elaganuru2024highly}. Nevertheless, no systematic study exists on the realization of cavity-QED conditions on the optical NCF. 

In this paper, we report the cavity-QED conditions using a CPCSC technique. The CPCSC is realized by combining the NCF and symmetric defect mode nano-grating (DMG). The maximum $\eta$-value up to $87$\% is realized by placing the SQE at the anti-node position of the CPCSC. The current platform paves the way for manipulating single photons with applications in quantum technology.

\section{Methodology}
\begin{figure}
	\includegraphics[width= 8 cm]{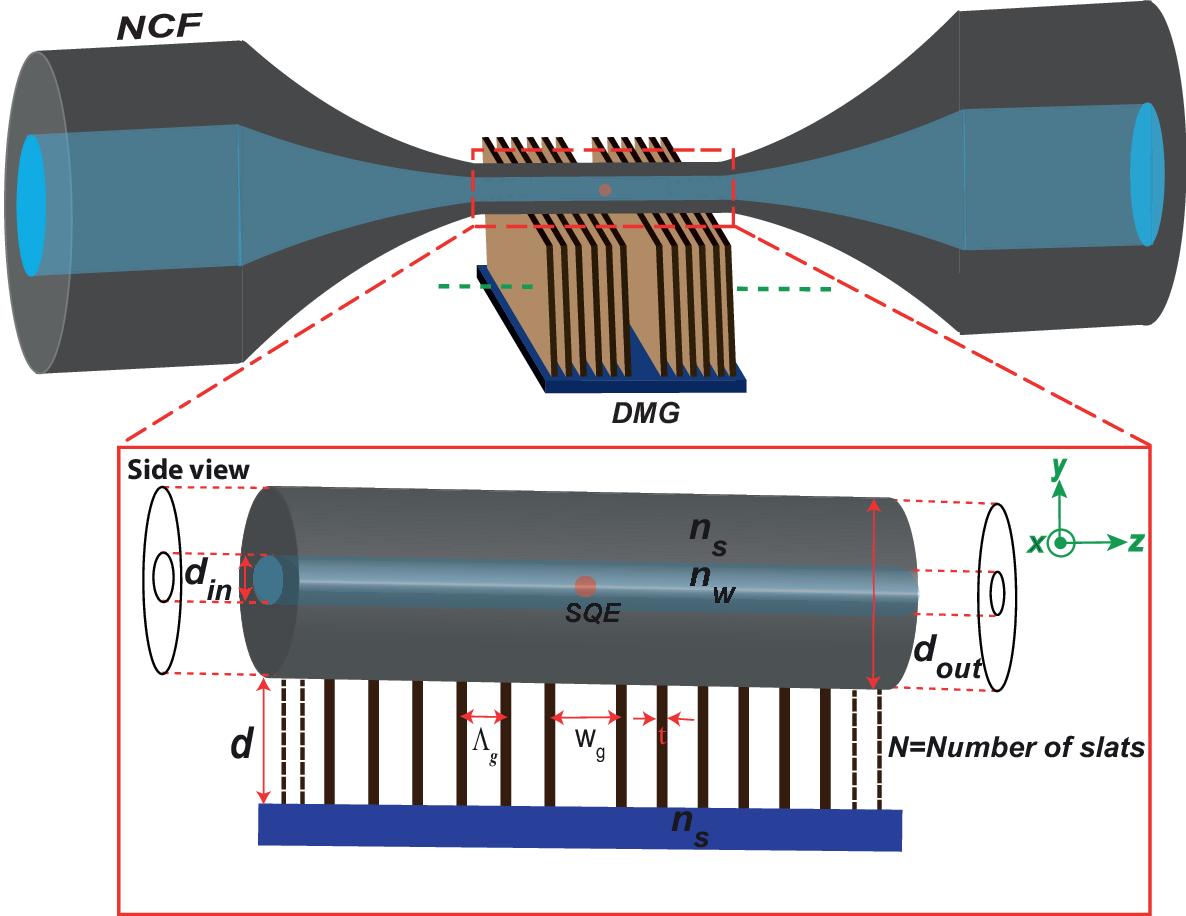}
	\caption{The conceptual diagram of composite photonic crystal symmetric cavity (CPCSC). The CPCSC is formed by combining a symmetric defect mode nano-grating (DMG) with the waist region of an optical nanocapillary fiber (NCF). The single quantum emitter (SQE) is placed at the anti-node position of the cavity. The inset shows a side view of the CPCSC. $d_{in}$ and $d_{out}$ are the subwavelength inner and outer diameters of the NCF, respectively. $n_w$ and $n_s$ are water and silica refractive indices, respectively. The DMG parameters are grating period, the width of defect, slat height, slat thickness, and number of slats denoted by $\Lambda{_g}$, $w_{g}$, $d$, $t$, and $N$, respectively.}\label{fig1}	
\end{figure}
Figure \ref{fig1} shows the conceptual sketch for realizing CPCSC on the optical NCF. The inset shows a side view and detailed parameters of the composite system. Here, the water and silica mediums are considered as inner ($d_{in}$) and outer ($d_{out}$) diameters of the NCF, respectively. Note that the present system is three-layered in contrast to the previous works \cite{nayak2018nanofiber}. The DMG has a rectangular-shaped grating pattern with grating period of ($\Lambda{_g}$) 244 nm and a slat height of $d$, extending from a silica substrate along $y$-direction. We fixed the defect width ($w_{g}$= $1.5 \Lambda_{g}$) and it is opened at the center of the DMG. The total width of the DMG (along $x$-direction) is treated as infinitely long. We design the cavity in such a way that the operating wavelength is around 620 nm to match the emission wavelength of a quantum emitter, such as semiconductor quantum dots and NV centers in nanodiamonds \cite{aharonovich2016solid, resmi2024channeling}.

We utilize the finite-difference time-domain method to perform all simulations in a fully three-dimensional manner using 3D-FDTD (Ansys Lumerical) software. The design consists of SQE, DMG, and NCF, is enclosed within a simulation region of 3$\times$4$\times$180 $\mu m^3$, with perfectly matched layers acting as walls to avoid the radiation reflections within the boundaries. The NCF is treated as infinitely long. We used an electric dipole as an SQE and a power monitor to determine the coupled power. In general, the channeling efficiency of spontaneous emission of the SQE through the NCF-guided modes in the presence of a cavity is defined as $\eta$= $\gamma_{c}/\gamma_{T}$, where $\gamma_{c}$ and $\gamma_{T}$ are the spontaneous emission rate coupled through the NCF-guided modes and total spontaneous emission rate of an SQE. In the present simulations, $\eta$ is defined as $\gamma_{c}/\gamma_{T}$= ${P_C}/{P_T}$, where $P_C$ is the power coupled through the NCF-guided modes from the SQE in the presence of the cavity, $P_T$ is the total power emitted by the SQE in the presence of the cavity.

Firstly, we optimize both the parameters of a DMG such as slat thickness ($t$) by monitoring duty cycle (${t}$= $\alpha \Lambda_{g}$) and total slat number ($N$) while maximizing the $\eta$-value through the NCF-guided modes. Subsequently, we optimize the NCF $d_{in}$ and $d_{out}$-values to maximize the $\eta$-value. Note that $d_{in}$ and $d_{out}$-values are in the subwavelength range. Note that $d$-value is fixed at 2 $\mu$m for all simulations \cite{sadgrove2013photonic}. We also study the cavity transmission ($T$) and reflection ($R$) spectra using $x$- and $y$-polarized input mode sources. The polarization response of the CPCSC system is investigated and systematically analyzed. Finally, we study the position dependency of the SQE in the cavity.
\section{RESULTS AND DISCUSSIONS}
\begin{figure}
	\includegraphics[width= 8 cm]{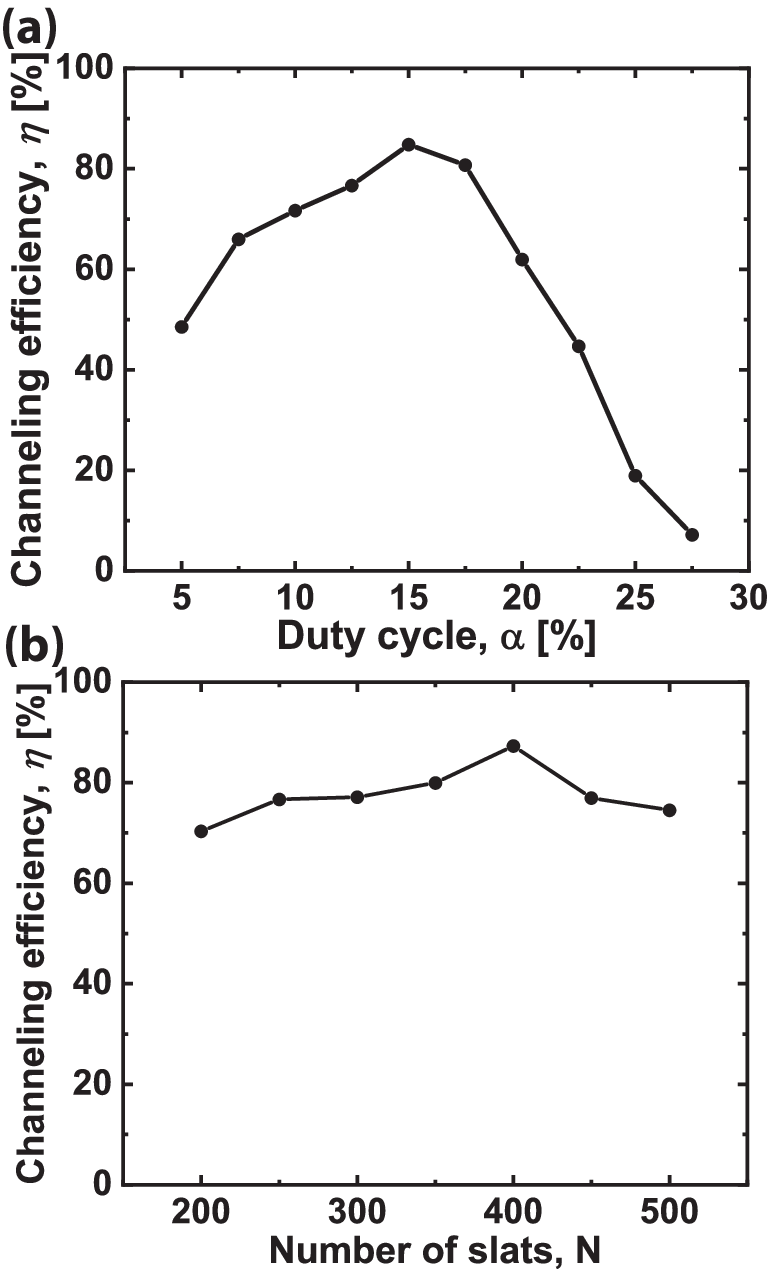}
	\caption{(\textbf{a}) and (\textbf{b}) Depict the dependency of channeling efficiency ($\eta$-value) on the duty cycle ($\alpha$) and number of slats ($N$), respectively.} \label{fig2}
\end{figure}
Figure \ref{fig2} (\textbf{a}) shows the simulation results for the variation of $\eta$-value with the $\alpha$-values. The horizontal axis represents the $\alpha$-value while the vertical axis represents the $\eta$-values. We optimize the $\alpha$-value, while maximizing the $\eta$-values. We sweep the $\alpha$-value from 5$\%$ to 27.5$\%$ with a step size of 2.5$\%$. The maximum $\eta$-value of 84$\%$ is obtained for 15$\%$, corresponding to a slat thickness ($t$) of 37 nm. Beyond this point, increasing the $\alpha$-value results in a decrease in $\eta$-value. This is due to increased scattering losses caused by the thicker slats. Note that we fixed the \textit{N}-value at 350 while optimizing the $\alpha$-value. Figure \ref{fig2} (\textbf{b}) shows the simulation results for the variation of $\eta$-value with the \textit{N}-values. The horizontal axis represents the \textit{N}-value while the vertical axis represents the $\eta$-values. We sweep the \textit{N}-value from 200 to 500 with a step size of 50. The maximum $\eta$-value of 87$\%$ is obtained for 400. Increase the $N$-value beyond this point leads to further decrease in $\eta$-value. This is can be attributed to the fact that scattering losses induced by the increased number of slats on both sides of a cavity. Here, we fixed the $\alpha$-value at 15$\%$ obtained from previous results while optimizing the \textit{N}-value. Note that we fixed the $d_{in}$, $d_{out}$, and $\Lambda{_g}$-values at 120 nm, 530 nm, and 244 nm, respectively while optimizing $\alpha$-value and \textit{N}-value.

\begin{figure}
	\includegraphics[width= 8 cm]{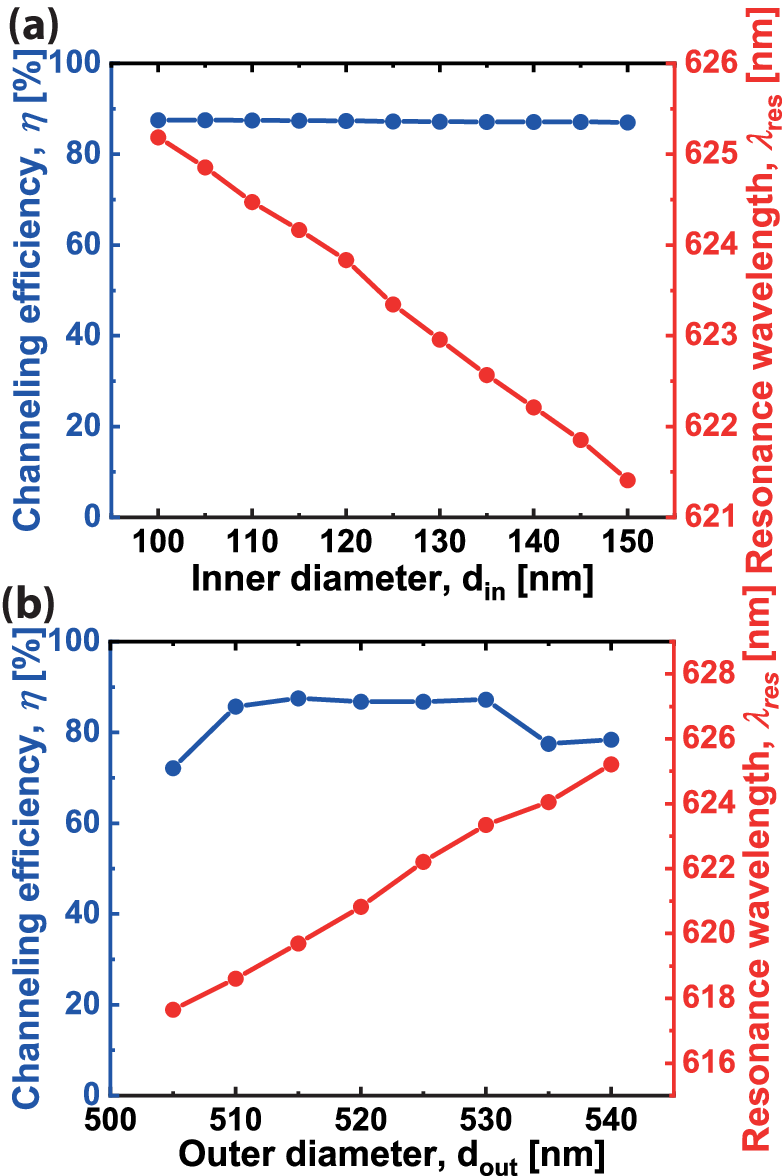}
	\caption{(\textbf{a}) and (\textbf{b}) Depict the dependency of channeling efficiency ($\eta$) (blue circles), cavity resonance wavelength ($\lambda_{res}$) (red circles) as a function of $d_{in}$-value and $d_{out}$-values, respectively.} \label{fig3}
\end{figure}

Figure \ref{fig3} (\textbf{a}) shows the simulation results for the dependency of $\eta$-value (blue circles) and $\lambda_{res}$-value (red circles) as a function of $d_{in}$-value of the NCF. The horizontal, left, and right vertical axes correspond to $d_{in}$, $\eta$, and  $\lambda_{res}$-values, respectively. We optimize the $d_{in}$-value, while maximizing the $\eta$-values. We sweep the $d_{in}$-value from 100 nm to 150 nm with a step size of 5 nm. The maximum $\eta$-value of 87$\%$ is realized. It is clear that $\eta$-value remains constant for different $d_{in}$-values. The optimized $d_{in}$-value is considered about 125 nm for experimental feasibility. One can also see that the $\lambda_{res}$-value decreases from 625 nm to 621 nm while increasing the $d_{in}$-value. This is because of the difference in the effective refractive indices between silica and water. It should be mentioned that we fixed the $d_{out}$-value at 530 nm while optimizing the $d_{in}$-value.

Figure \ref{fig3} (\textbf{b}) shows the simulation results for the variation of $\eta$-value (blue circles) and $\lambda_{res}$-value (red circles) with the $d_{out}$-values of the NCF. The horizontal, left, and right vertical axes correspond to $d_{out}$, $\eta$, and  $\lambda_{res}$-value, respectively. We optimize the $d_{out}$-value, while maximizing the $\eta$-values. We fix the $d_{in}$-value at 125 nm obtained from previous results while optimizing the $d_{out}$-value. We sweep the $d_{out}$-value from 505 nm to 540 nm with a step size of 5 nm. The maximum $\eta$-value of 87.5$\%$ is obtained for 515 nm. The maximum and minimum $\lambda_{res}$-values are 628 nm and 616 nm, respectively. One can also see that the tunability of cavity $\lambda_{res}$-value around the designed wavelength of 620 nm \cite{yalla2020design}. Therefore, the optimized parameters for maximum $\eta$-value are $\Lambda{_g}$= 244 nm, $\alpha$= 15$\%$, \textit{N}= 400, $d_{in}$= 125 nm, and $d_{out}$= 515 nm.  

 \begin{figure}
	\includegraphics[width= 8 cm]{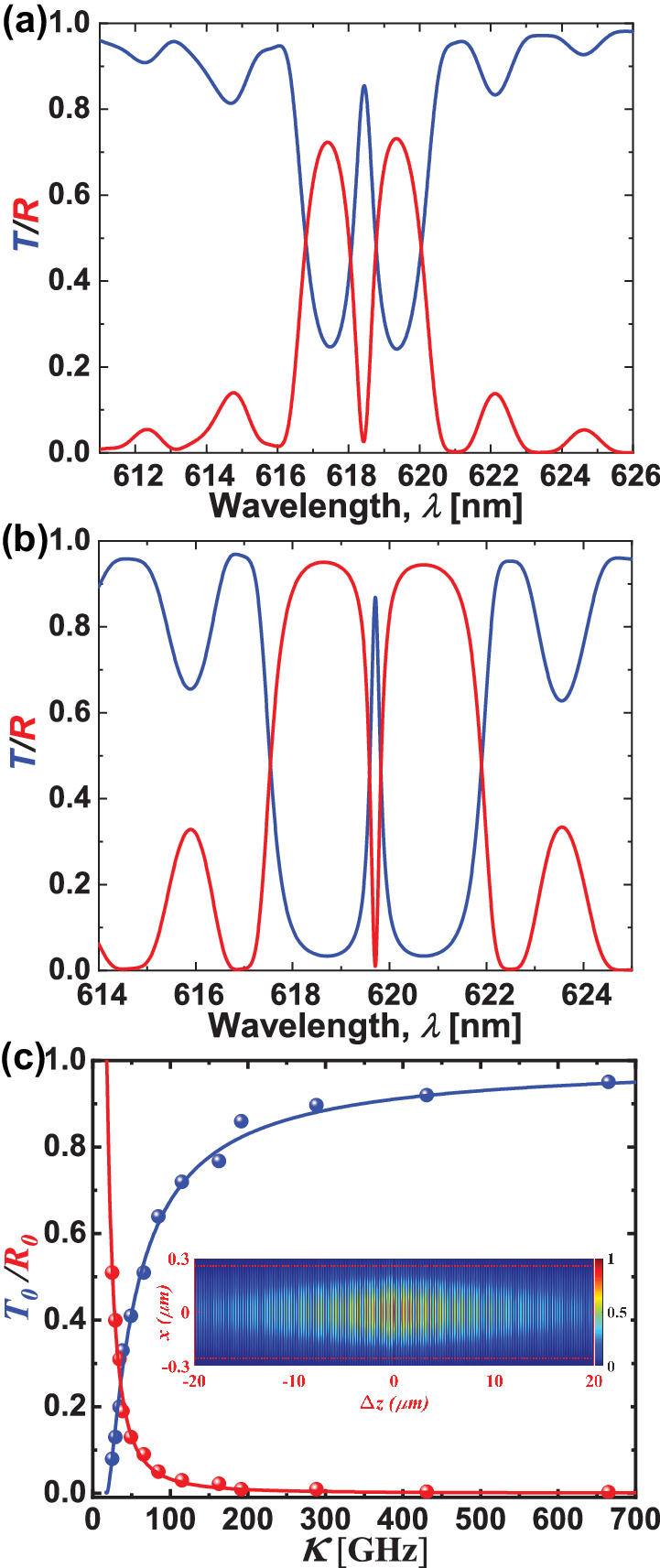}
	\caption{(\textbf{a}) and (\textbf{b}) Depict the cavity transmission ($T$) (blue trace)/reflection ($R$) (red trace) spectra for $x$- and $y$-polarized input mode source, respectively. (\textbf{c}) shows the variation of on-resonance transmissivity ($T_{0}$, blue spheres) and reflectivity ($R_{0}$, red spheres) as a function of cavity field decay rates ($\kappa$). The solid blue and red traces represent fits to the data. The inset shows the normalized electric field intensity distribution profile in $xz$-plane at $y$= 0 for $y$-polarized SQE placed at the cavity anti-node ($\Delta{z}$= 0 nm). The red dotted lines represent the NCF surfaces.} \label{fig4}
\end{figure}
\begin{figure*}
	\includegraphics[width=\linewidth]{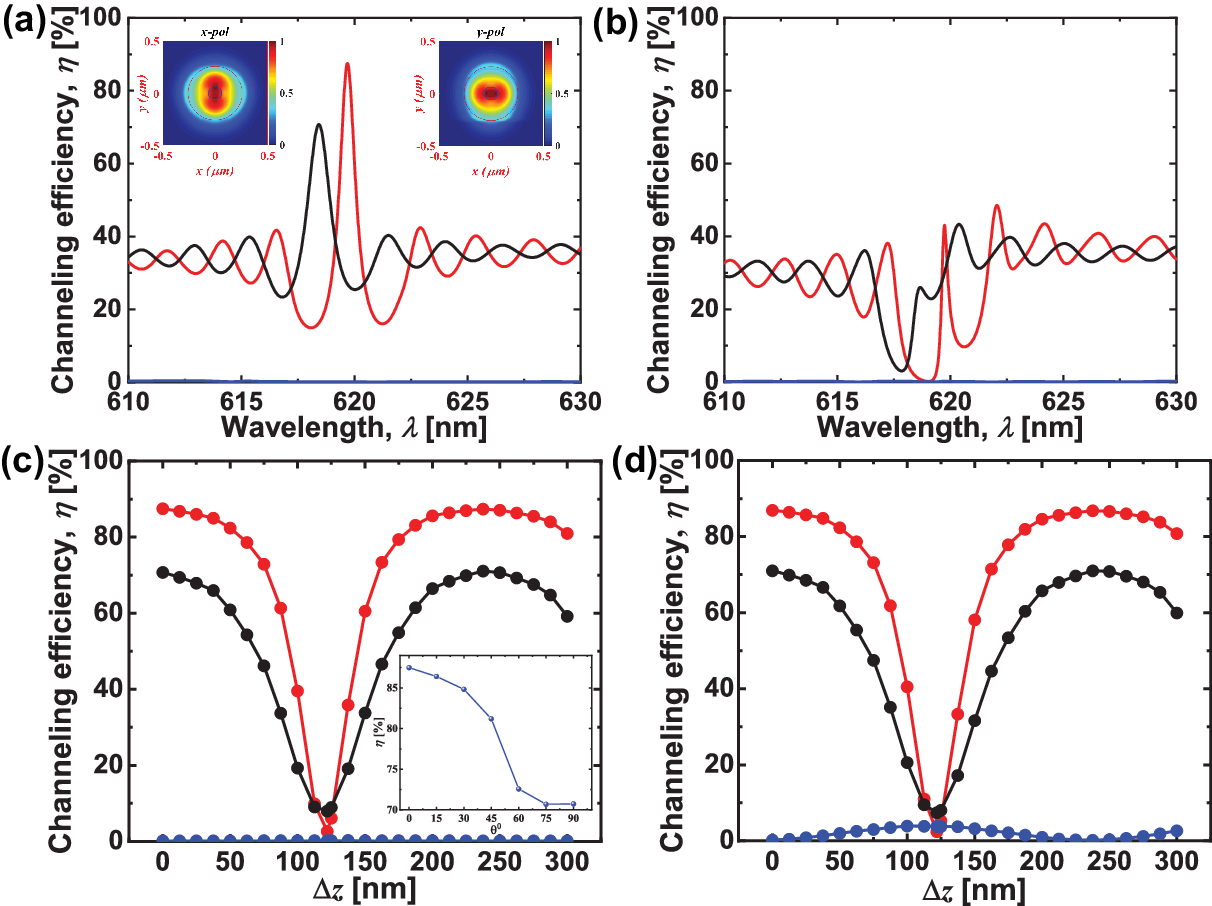}
	\caption{(\textbf{a}) and (\textbf{b}) Depict the variation of channeling efficiency ($\eta$-value) for three polarizations i.e, $x$ (black), $y$ (red) and $z$ (blue) of SQE when it is placed at anti-node ($\Delta{z}$= 0 nm) and node ($\Delta{z}$= 100 nm) of the cavity, respectively. The insets show the simulated transverse spatial mode profiles of the cavity determined at $\Delta{z}$= 183 nm. The left and right correspond to the $x$ and $y$-modes of the cavity, respectively. The blue and red circles denote the inner and outer diameters of the NCF. (\textbf{c}) and (\textbf{d}) show the variation of $\eta$-value with the $\Delta{z}$-values for three polarizations i.e., $x$ (black), $y$ (red) and $z$ (blue) of SQE when it is positioned at the center (0 nm) and off-center (50 nm) of the NCF. The inset shows the dependency of $\eta$-value on the orientation of the SQE in $xy$-plane, where $\theta$= $0^0 (90^0)$ corresponds to $y$ ($x$)-pol.} \label{fig5}
\end{figure*}
Figures \ref{fig4} (\textbf{a}) and (\textbf{b}) show the transmission ($T$) (blue trace) and reflection ($R$) (red trace) spectra for $x$ ($x$-pol) and $y$ ($y$-pol) polarized mode sources, respectively. The horizontal axis corresponds to wavelength (\textit{$\lambda$}) while the vertical axis corresponds to \textit{$T/R$}. It shows that a strong photonic stop band is formed between 617/618 nm and 620/622 nm for the $x$/$y$-pol of the cavity with a transmission peak at the center of the stop band, respectively. The $\lambda_{res}$ and ${\Delta}\lambda$-values are 618 (619) nm and 0.655 (0.248) nm for $x$ ($y$)-pol, respectively. The corresponding quality factor ($Q$) value is 944 (2498). It is clear that, $Q$-value is higher for the $y$-pol, which can be attributed to the higher reflectivity induced by the DMG along the $y$-direction compared to the $x$-direction \cite{yalla2014cavity}. 

Furthermore, we analyze the on-resonance cavity's  transmissivity ($T_{0}$) and reflectivity  ($R_{0}$) by varying $N$-values. The corresponding cavity field decay rates, $\kappa$ (= $\frac{c \Delta \lambda}{\lambda_{res}}$, where \textit{c} is the speed of light in free space) values are determined. The summary of the results is shown in Fig. \ref{fig4} (\textbf{c}). The horizontal axis represents the $\kappa$-values and the vertical axis corresponds to the respective $T_{0}$/$R_{0}$-values. Blue and red spheres are for the $T_{0}$ and $R_{0}$-values, respectively.  Note that the $\kappa$-value decreases while increasing the $N$-value. One can readily see $T_{0}$/$R_{0}$-value increase/decreases with increase of the $\kappa$-value. It is clear that the on-resonance $T_{0}$-value is more than 70\% even for the cavity modes having $\kappa$-values in the range of 110 to 430 GHz, and the $\eta$-value (see Fig. \ref{fig2} (\textbf{b})) exceeds 70\%. To extract $\kappa_{sc}$-value, we followed the analytical formulism for a symmetric cavity as described in Ref. \cite{keloth2017fabrication}. The on-resonance transmissivity and reflectivity are		
$T_{0}$= $\bigm|1-\dfrac{\kappa_{sc}}{\kappa}\bigm|^2$  and $R_{0}$= $\bigm|\dfrac{\kappa_{sc}}{\kappa}\bigm|^2$, respectively. $\kappa$ and $\kappa_{sc}$ are the total cavity field decay rate and scattering rate (intra-cavity loss rate), respectively. It is clear that, both $T_{0}$ and $R_{0}$-values become equal when $2 \kappa_{sc}$= $\kappa$. To find the scattering-limited decay rate of the cavity ($\kappa_{sc}$), we fit the above equations (blue and red traces) with $\kappa_{sc}$ as a free parameter. The determined $\kappa_{sc}$-value is 18 GHz. Additionally, we examined the behavior of the cavity for $x$-polarized input. We found that higher $T_{0}$-values and lower $\kappa_{sc}$-value, which implies better performance of the cavity. This is due to the fact that the scattering loss is lower for $x$-pol compared to $y$-pol. To extract cavity effective length ($l_{eff}$), we simulated the cavity mode intensity profile and shown in the inset of Fig. \ref{fig4} (\textbf{c}). It shows the normalized electric field intensity distribution in $xz$-plane at $y$= 0 for $y$-polarized SQE placed at cavity anti-node ($\Delta{z}$= 0 nm). The red dotted lines represent the NCF surfaces. One can readily see that the intensity decays exponentially from the center. We find the $l_{eff}$-value to be $\approx$ 28 $\mu m$. Using the extracted $l_{eff}$ and $\kappa_{sc}$-values, we determine the cavity performance parameters such as scattering limited quality factor ($Q_{sc}$), cavity finesse ($\mathcal{F}_{sc}$), and one-pass power loss ($\mathcal{L}$). We found $Q_{sc}$, $\mathcal{F}_{sc}$, and $\mathcal{L}$-values to be 26894, 297 and 1.04\%, respectively \cite{yalla2022one}. 

The present CPCSC system operates effectively in the ``Purcell regime" of cavity QED as the $\kappa$-values are in the GHz range. In this regime, the purcell factor ($F_{P}$) can be approximately equal to the single QE-cavity cooperativity ($C$). Therefore, $F_{P}\approx C$= $\frac{(2g_{0})^2}{\kappa \gamma}$, where $2g_0$ is the single-photon Rabi frequency (SQE–cavity coupling rate) and $\gamma$ is the spontaneous emission decay rate of the SQE \cite{yalla2014cavity, keloth2017fabrication}. At the optimum $N$-value of 400, we found the $F_{P}$ and $\kappa$-values are 11 and 193 GHz, respectively. Considering $\gamma$= 1.2 GHz, which is a typical value for NV centers in nanodiamond \cite{zhao2013observation}, we estimate the $2g_{0}$-value to be 50 GHz. The present cavity system can be effectively integrated with other solid-state emitters such as SiV centers in nanodiamond and semiconductor quantum dots \cite{aharonovich2016solid, resmi2024channeling}. Using the optimum $\kappa$ and $l_{eff}$-values, we find cavity finesse ($F$) to be $\approx$ 28. The maximum expected $\eta$-value of $\sim$94\% can be achieved through the NCF-guided modes \cite{le2009cavity}. On the other hand, using the extracted $C$-value, we have analytically estimated a cavity-enhanced $\eta$-value of $\sim$93\% \cite{nayak2019real}. However, both formalisms assume no scattering losses from the cavity mirrors. In practical systems, finite losses are induced by the DMG, which limit the maximum $\eta$-value to 87\%. To realize higher $\eta$-values compared to the present case, one approach is to implement the apodized refractive index variations within the cavity design \cite{nayak2013photonic, quan2010photonic}. One can minimize the grating-induced scattering losses and thereby enhance the $\eta$-value.

Figures \ref{fig5} (\textbf{a}) and (\textbf{b}) show the dependency of $\eta$-value as a function of $\lambda$-values for $x$- (black), $y$- (red) and $z$- (blue) polarized SQE, when it is placed at anti-node ($\Delta{z}$= 0 nm) and node ($\Delta{z}$= 100 nm) position of the cavity, respectively. In Fig. \ref{fig5} (\textbf{a}), it is clear that the maximum $\eta$-value of $87\%$ ($71\%$) is obtained for $y$ ($x$)-pol at a wavelength of 619 (618) nm. It is also clear that the maximum $\eta$-value obtained $y$-pol is as expected. This is due to the NCF-guided modes experiencing more refractive index modulation by the DMG along $y$-direction to the cavity, and SQE is positioned at the cavity's anti-node position ($\Delta{z}$= 0 nm). The insets show the simulated cavity transverse spatial mode profiles determined at $\Delta{z}$= 183 nm i.e. on the slat. The left and right correspond to the $x$ and $y$-modes of the cavity, respectively. The blue and red circles denote the inner and outer diameters of the NCF, respectively. One can readily see that the spatial mode profile of the $y$-mode is altered along the negative $y$-direction. This asymmetry is attributed to the presence of the DMG introduced along the $y$-direction. In Fig. \ref{fig5} (\textbf{b}), the minimum $\eta$-value of $1\%$ (3$\%$) for $y$ ($x$)-polarization occurs at a wavelength of 619 (618) nm. In this case, the SQE is positioned away from the anti-node position ($\Delta{z}$= 100 nm) of the cavity. The $\eta$-value for the $z$-polarization is less than $2\%$ for both cases. This is due to weak coupling between $z$-polarized SQE with the cavity's modes \cite{le2009cavity}.  

To deeper understand the coupling mechanism and robustness of the system, we further investigate the dependency of the $\eta$-value on the position and orientation of SQE. Figures \ref{fig5} (\textbf{c}) and (\textbf{d}) depict the variation of $\eta$-value with the $\Delta{z}$-values for $x$- (black), $y$- (red) and $z$- (blue) polarized SQE, when it is positioned at the center (0 nm) and off-center (50 nm) of the NCF, respectively. It is clear that the $\eta$-value of more than $85\%$ ($71\%$) for $y$ ($x$)-polarization at $\Delta{z}$-value of 0. This is due to the SQE being positioned at the anti-node positions of the cavity. It is also clear that the $\eta$-value decreasing from $85\%$ ($71\%$) to $2\%$ ($7\%$) for $y$ ($x$)-polarization at $\Delta{z}$-values from 0 nm to 122 nm. This is due to the SQE being positioned away from the anti-node position of the cavity. The $\eta$-value for the $z$-polarization, when the SQE is at the off-center (50 nm) of the cavity, is more ($4\%$) compared to when the SQE is at the center (0 nm) of the NCF. This is due to the SQE being positioned near the cavity. One can readily see that $\eta$-value oscillates between maxima (anti-node) and minima (node) with a period of $\sim$120 nm. This is attributed to the fact that the periodic oscillating electric field intensity of the cavity as seen in the inset of Fig. \ref{fig4} (\textbf{c}). It is also evident that $\eta$-value exhibits broad maxima spanning $\sim$100 nm, consistent with the behavior as reported in Refs. \cite{le2009cavity, li2018tailoring, sahu2022optimization}. From Fig. \ref{fig5} (\textbf{a}), it is also evident that the $\eta$-value strongly depends on the polarization orientation of the SQE in the $xy$-plane. Therefore, we simulated $\eta$-value for different orientations of the SQE in the $xy$-plane. The summary of the results is shown in the inset of Fig. \ref{fig5} (\textbf{c}), where $\theta$= $0^0 (90^0)$ corresponds to $y$ ($x$)-polarization. It is clear that the $\eta$-values lie between the maximum $y$ and $x$-pol. From the experimental perspective, precise control over the dipole orientation remains challenging. However, a notable advantage of employing a composite cavity case is its ability to adjust the position of the DMG relative to the NCF, as well as the alignment of the SQE in the $xy$-plane. This flexibility facilitates optimization of LMI i.e. $\eta$-value, even when the dipole orientation is not pre-aligned along a specific direction in the $xy$-plane. We already examined the robustness of the present design by systematically studying the dependency of $\eta$-value on several parameters, including the slat number, the inner and outer diameters of the NCF, and the polarization, orientation, and position of the SQE.

Three key challenges regarding the experimental feasibility of the present simulations are to be addressed. The first is the fabrication of the NCF with such a small capillary hole; the second is the formation of the cavity on the NCF; the third is placing the SQE at the anti-node of the cavity. Regarding the NCF, commercial capillary optical fibers can be tapered to sub-wavelength diameters to realize NCF using a heat and pull technique \cite{bashaiah2024fabrication}. As seen in Fig. \ref{fig3}, one can see that the peak $\eta$-values occurred somewhat broadly with respect to the inner and outer diameters of the NCF. This suggests that the $\eta$-value is not significantly impacted by even a small amount of uncertainty in the inner and outer diameters of the NCF. As a result, the experimental realization would have more freedom. Regarding the formation of the cavity on the NCF, the cavity formation on the waist region of ONF was demonstrated by combining the external DMG with the ONF  \cite{sadgrove2013photonic, yalla2014cavity, keloth2015diameter, yalla2022one}. The maximum $\eta$-value of $65$\% was demonstrated through the guided modes of SMF \cite{yalla2014cavity}. Additionally, the one-sided cavity, in which DMG is asymmetric about the defect center was demonstrated \cite{yalla2022one}. Therefore, one can realize the present cavity on the NCF using the composite technique. Regarding the positioning of the SQE at the cavity anti-node, one can achieve this in two ways. One is that the SQE is introduced into the NCF by injecting water containing dissolved quantum dots serving as a single light emitter using either capillary force \cite{white2006liquid} or pump technique \cite{liu2013hollow} through one end of such a thin NCF. Then, the cavity is formed by integrating the DMG such that the SQE coincides with the cavity anti-node. The other is to form a composite cavity onto the waist region of the NCF, then inject the water with dissolved quantum dots using above mentioned techniques while slowly monitoring the flow rate to ensure an SQE is placed at the anti-node of the cavity. In both scenarios, one can realize the accurate positioning of an SQE at the desired cavity anti-node. It is worth noting that the same spontaneous emission enhancement can also be achieved by placing the SQE at any of the cavity anti-nodes within the uniform region as seen in the inset of Fig. \ref{fig4} (\textbf{c}) \cite{yalla2014cavity}. The presence of water surrounding the SQE may affect its spectral properties, such as spectral broadening and spatial diffusion \cite{gupta2013spontaneous}, which can affect the SQE-cavity coupling strength ($2g_{0}$). Nevertheless, such effects can be mitigated by choosing a suitable SQE and working at cryogenic temperatures. The promising candidates are NV/SiV centers in nanodiamond and high quantum efficiency semiconductor quantum dots \cite{aharonovich2016solid}.   

One can readily see that the $\eta$-value is on par with other ONF-cavities \cite{nayak2019real,sahu2022optimization}. However, $\eta$-value can be further enhanced by increasing the refractive index of the medium and NCF/DMG. Simulations indicate that if silica is replaced with another high refractive index material, such as diamond, silicon nitride, or gallium phosphate, the $\eta$-value may exceed the present value \cite{hausmann2013coupling, das2023efficient, resmi2024highly}. Additionally, if we substitute water with any other higher refractive index liquid, the $\eta$-value may be impacted \cite{faez2014coherent}. Note that one can channel the total spontaneous emission rate of the SQE into one-sided guided modes of ONF using the asymmetric cavity \cite{yalla2022one}, which finds applications in quantum technology. 

\section{Conclusion}
We reported the cavity QED conditions using a composite photonic crystal symmetric cavity (CPCSC) on an optical NCF. The CPCSC is formed by combining the NCF and DMG. The maximum $\eta$-value upto $87$\% is realized by placing the SQE at the anti-node position of the CPCSC. The current platform paves the way for manipulating single photons with applications in quantum technology.

\begin{acknowledgments}
RRY acknowledges financial support from the Scheme for Transformational and Advanced Research in Sciences (STARS) grant from the Indian Institute of Science (IISc), Ministry of Human Resource Development (MHRD) (File No. STARS/APR2019/PS/271/FS) and Institute of Eminence (IoE) grant at the University of Hyderabad, Ministry of Education (MoE) (File No. UoH-IoE-RC2-21-019). Science and Engineering Research Board (SERB) for the Core Research Grant (CRG) (File No. CRG/2021/009185), the Council of Scientific and Industrial Research (CSIR) from the Human Resources Development Group (HRDG) (File No. 03/1487/2023/EMR-II).
\end{acknowledgments}

\section*{DATA AVAILABILITY}
The data that support the findings of this study is available from the corresponding author upon reasonable request.

%\section*{References}
\bibliography{apssamp}% Produces the bibliography via BibTeX.

\end{document}